\documentclass{appolb}
\usepackage{graphicx}

\usepackage[comma,square,numbers,sort&compress]{natbib}
\usepackage{amsmath}
\usepackage{amssymb}
\usepackage{microtype}

\newcommand{\pt}{\ensuremath{p_{\rm T}}\xspace}

\begin{document}
\title{Event-activity dependence of heavy-flavor production at the ALICE experiment
\thanks{Presented at the V4-HEP Theory and Experiment in High Energy Physics workshop, Prague, Czechia (2024)}%
}
\author{Róbert Vértesi
\address{HUN-REN Wigner Research Centre for Physics \\ 
Konkoly-Thege Miklós út 29-33, 1121 Budapest, Hungary}
\\[3mm]
{(for the ALICE Collaboration)
}
}
\maketitle
\begin{abstract}
Heavy-flavor production at the LHC offers valuable tests of quantum-chromodynamics calculations, owing to the large masses of heavy quarks. Measurements of charm production as a function of event activity reveal new features of charm production and fragmentation, providing insights to the interplay between soft and hard processes. In addition, charm production in heavy-ion collisions addresses flavor-dependent quark transport properties in both hot and cold nuclear matter, helping to clarify the roles of coalescence and fragmentation in heavy-flavor hadron formation. 
This contribution summarizes recent measurements from the ALICE experiment on charm production as a function of charged-particle multiplicity in pp collisions at various energies, including the measurements of charm baryon-to-meson production yield ratios in pp, p--Pb and Pb--Pb collisions. New results on ${\rm D}^0$ production in pp collisions as a function of the transverse spherocity of the event, as well as of the transverse event-activity classifier $R_{\rm T}$, are also presented.
\end{abstract}
  
\section{Introduction}
Small collision systems (such as pp and p--Pb) with high final-state multiplicity exhibit similar signatures to those observed in heavy-ion collisions, where the formation of quark--gluon plasma
(QGP) is expected. These signatures include strangeness enhancement and long-range multiparticle correlations (collectivity, often called ``flow'')~\cite{ALICE:2016fzo,ALICE:2019zfl}.
The question naturally emerges whether quark--gluon plasma can be produced in such systems, or if there are vacuum-QCD effects that are responsible for this behavior.

Heavy-flavor (HF) quarks work as hard probes down to low transverse momentum (\pt), thus they can be used for pQCD benchmark measurements. Measuring the dependence of heavy-flavor production on charged-hadron
multiplicity and event activity allows for the investigation of collective-like effects from small to large systems, the interplay between the hard and soft particle production, the role of multiparton interactions in heavy-quark production, and charm fragmentation across different collision systems.

The ALICE experiment is described in detail in Ref.~\cite{ALICE:2022wpn}. The central barrel is in a magnetic field and covers the pseudorapidity range $|\eta|<0.9$. The Inner Tracking System (ITS) and the Time Projection Chamber (TPC) are used for charged-particle tracking. The ITS is responsible for the reconstruction of primary (collision) and secondary (decay) vertices, while the TPC can identify charged particles based on specific energy loss. The V0 detector, with a pseudorapidity coverage of $-3.7<\eta<-1.7$ and $2.8<\eta<5.1$, is used for event characterization. Photon and electron identification is facilitated by the Electromagnetic Calorimeter. There is a Muon Spectrometer in the forward direction $-4<\eta<-2.5$, that provides muon triggers and reconstructs their trajectories. 
Heavy-flavor hadrons can be reconstructed either from fully hadronic final states, or by observing leptons from semileptonic decay channels. Heavy-flavor identification is aided by secondary vertex reconstruction from charged-particle trajectories recorded by the ITS.

\section{Results}

The ALICE collaboration has a broad heavy-flavor program. Only a fraction of recent preliminary results can be highlighted in the current contribution.

\subsection{Event-activity-dependent production in pp collisions}

\begin{figure}[htb]
	\centerline{%
		\includegraphics[width=.85\columnwidth]{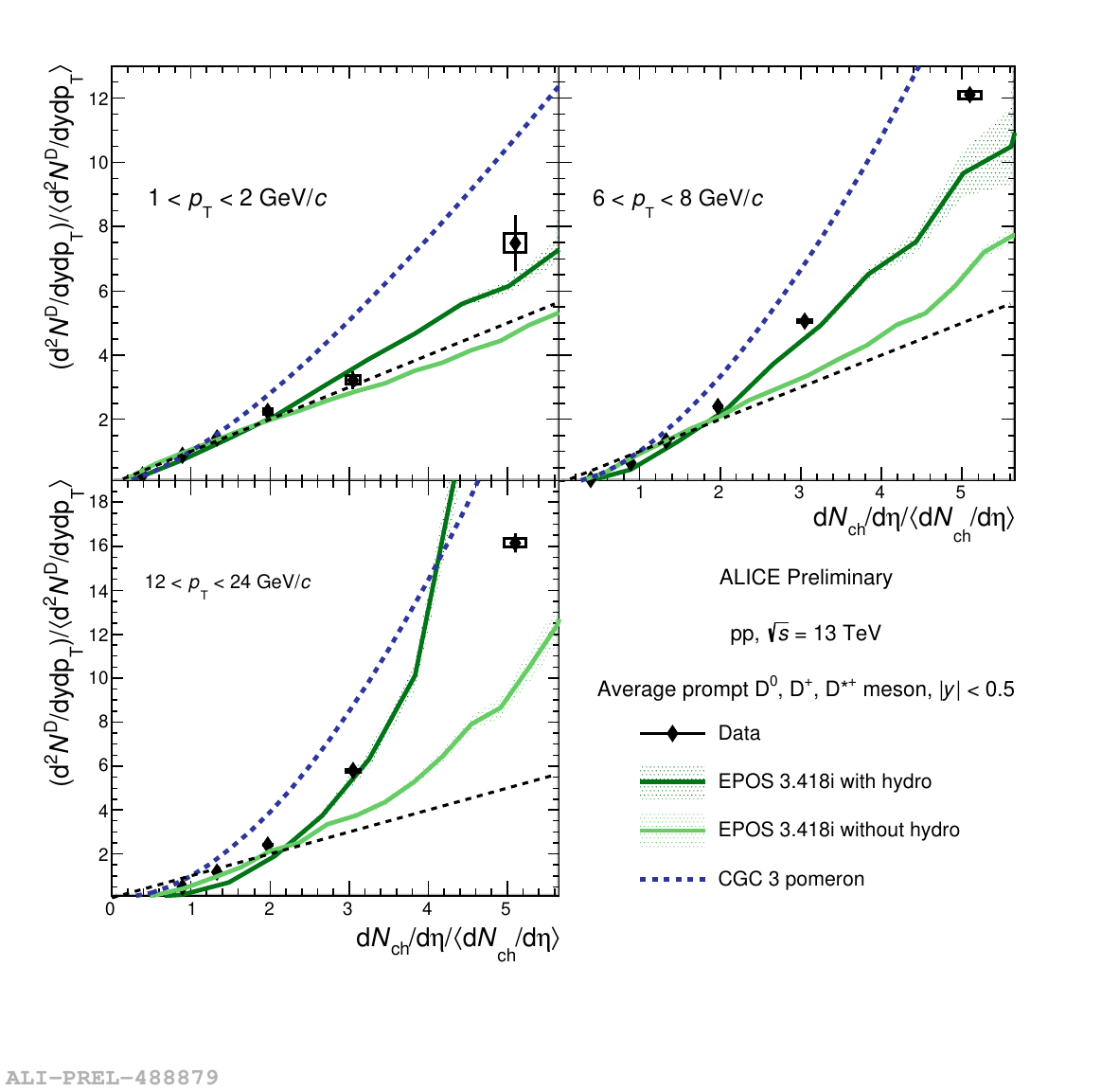}}
	\caption{The average prompt ${\rm D}^0$, ${\rm D}^+$ and ${\rm D}^{*+}$ self normalised yields as a function relative charged-particle multiplicity in pp collisions at $\sqrt{s}=13$ TeV at central rapidity, compared with EPOS calculations.}
	\label{fig:DvsMult}
\end{figure}

\begin{figure}[htb]
	\centerline{%
		\includegraphics[width=.475\columnwidth]{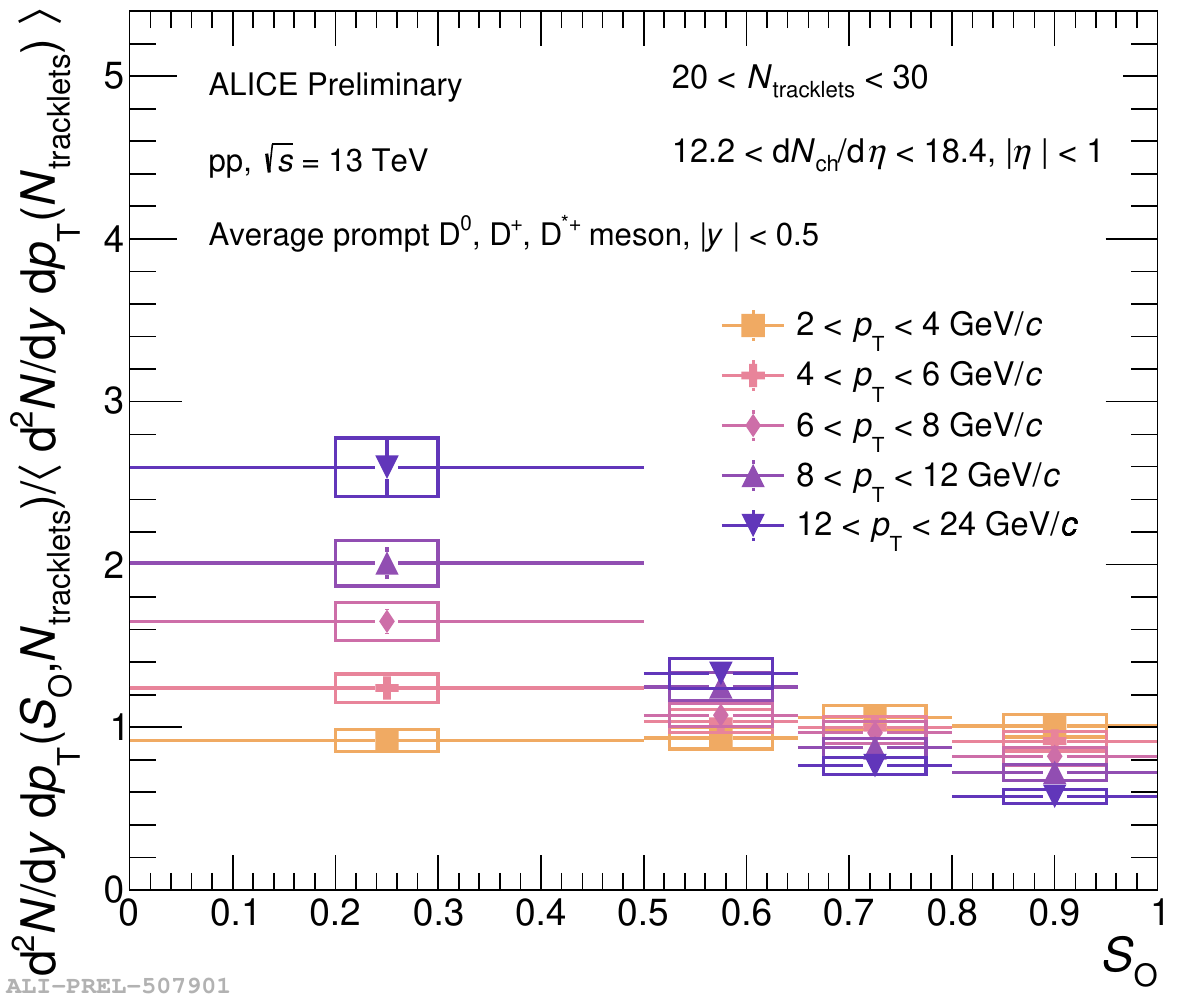}\hfill%
		\includegraphics[width=.475\columnwidth]{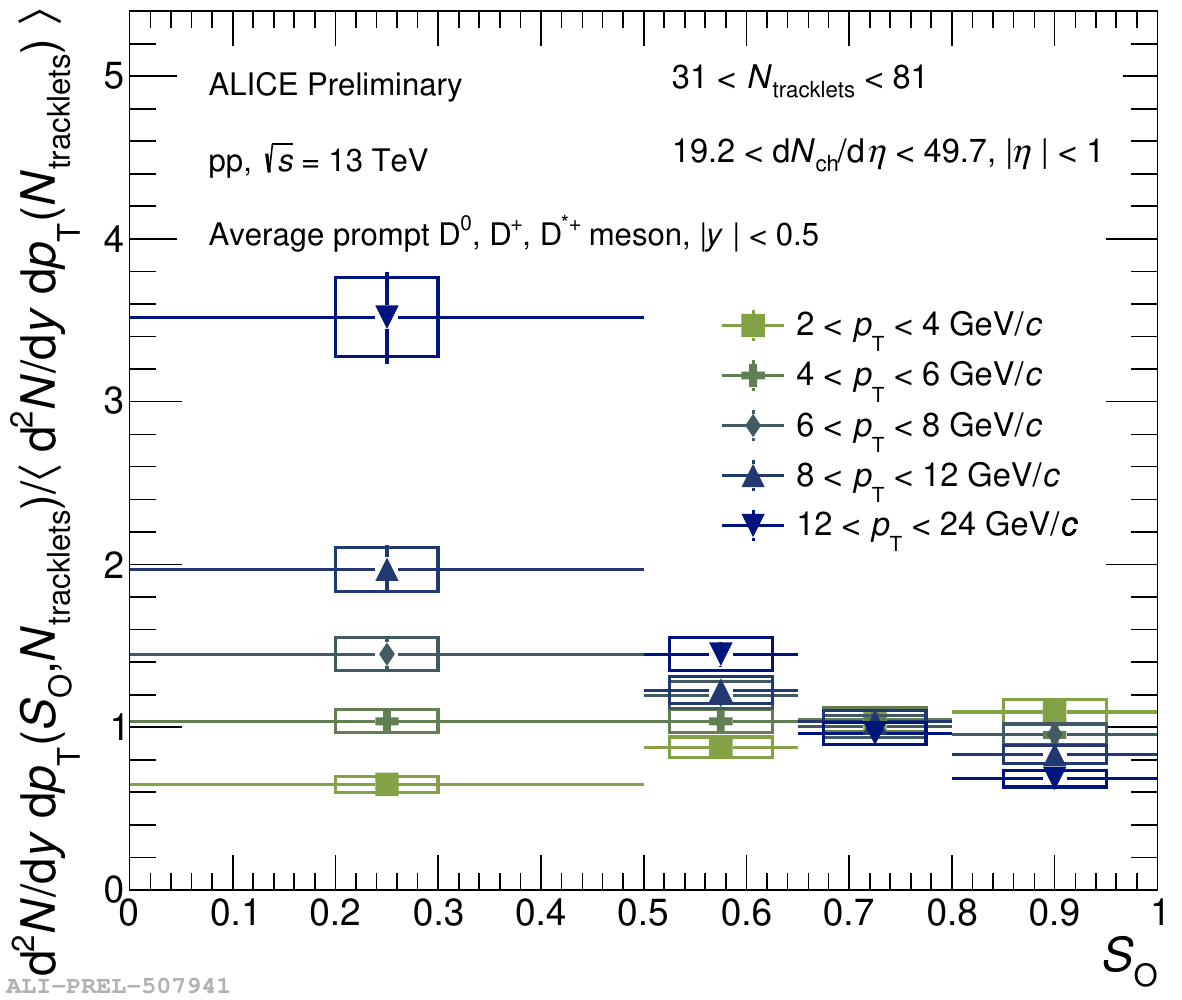}}
	\caption{
	Average D-meson self-normalized yield as a function of spherocity in pp collisions at $\sqrt{s}=13$ TeV in different \pt intervals for low (left panel) and high (right panel) multiplicity ranges.
	}

	\label{fig:DvsSphero}
\end{figure}

The measurement of the self-normalized heavy-flavor hadron yields with respect to final-state particle multiplicity provides strong constraints for models. 
Figure~\ref{fig:DvsMult} shows the results for D mesons at $\sqrt{s} = 13$ TeV. 
The EPOS parton model with hydrodynamic evolution~\cite{Pierog:2013ria} captures the trends seen in data correctly. Note that PYTHIA~8 with multiple-parton interactions is capable of describing self-normalized heavy-flavor electron production properly~\cite{ALICE:2023xiu}. 
Since the multiplicity-dependent production of HF hadrons is sensitive to auto-correlation, it requires a good simultaneous description of jets and the underlying event (UE). This difficulty can be mitigated by using multi-differential observables. One way to distinguish the soft and hard components is to use the transverse spherocity~\cite{Ortiz:2015ttf}, an event-shape observable that approaches zero for jetty events and unity for events that are isotropic in the plane transverse to the beam axis. Figure~\ref{fig:DvsSphero} shows the spherocity-dependent D-meson production in several transverse-momentum bins in the low and high multiplicity classes.
The results show a hint of an enhanced D-meson production toward higher multiplicity in jetty events. This suggests that the effect of hard scatterings lead to an increase in charged-particle multiplicity on the average. Note that autocorrelations may also play a role in forming the steeper-than-linear trend.


\subsection{Fragmentation in pp collisions}

Heavy-flavor production is usually described using the factorization hypothesis, that is, the production cross section of heavy-flavor hadrons can be expressed as the convolution of three independent factors: the parton distribution functions (PDF) of the incoming hadrons, the hard parton scattering cross section, and the fragmentation of the outgoing partons into heavy-flavor hadrons.
Charm-quark fragmentation fractions into different hadrons $f({\rm c}\rightarrow {\rm h}_{\rm c})$ show a reduction of ${\rm D}^0$-meson fractions by about 1/3, and an enhancement of charmed baryons in pp collisions both at $\sqrt{s}=5.02$ and 13 TeV, compared to ${\rm e}^+ {\rm e}^-$ collisions at HERA and LEP~\cite{ALICE:2023sgl}. The $Lambda_{\rm c}^+/{\rm D}^0$ ratio indicates a significant excess in pp compared to ${\rm e}^+ {\rm e}^-$ collisions toward low \pt values.
This shows that hadronization is not universal across different systems. Several models, based on the introduction of color junctions, statistical hadronization with augmented charm states and charm--light quark coalescence, succesfully describe the \pt-dependent trends of enhanced $\Lambda_{\rm c}^+/{\rm D}^0$~\cite{He:2019tik,Christiansen:2015yqa,Plumari:2017ntm}. In a recent ALICE publication~\cite{ALICE:2021npz} it was shown that the enhancement depends also on charged-hadron final-state multiplicity. 

\begin{figure}[htb]
	\centerline{%
		\includegraphics[width=.42\columnwidth]{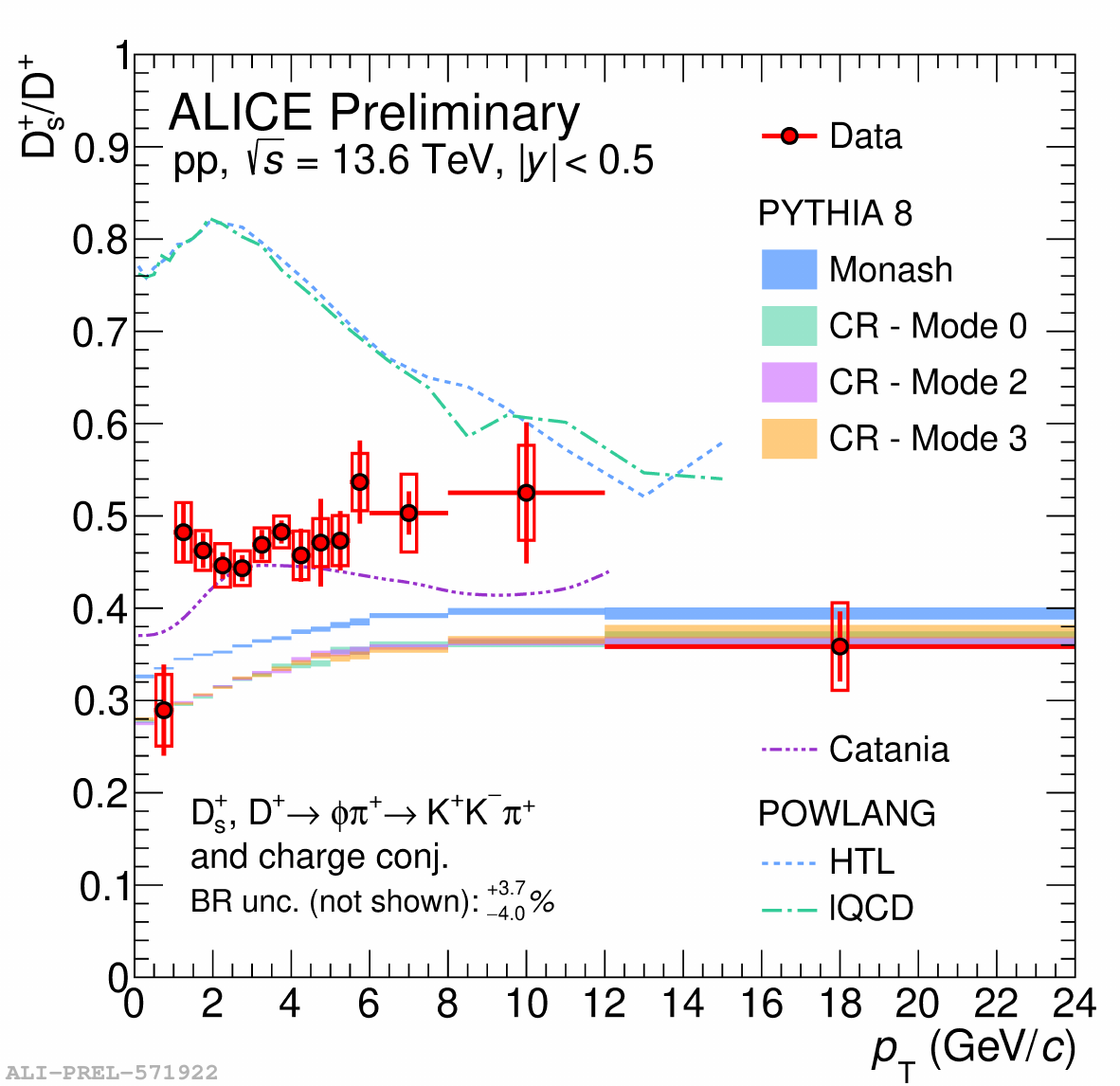}%
		\includegraphics[width=.58\columnwidth]{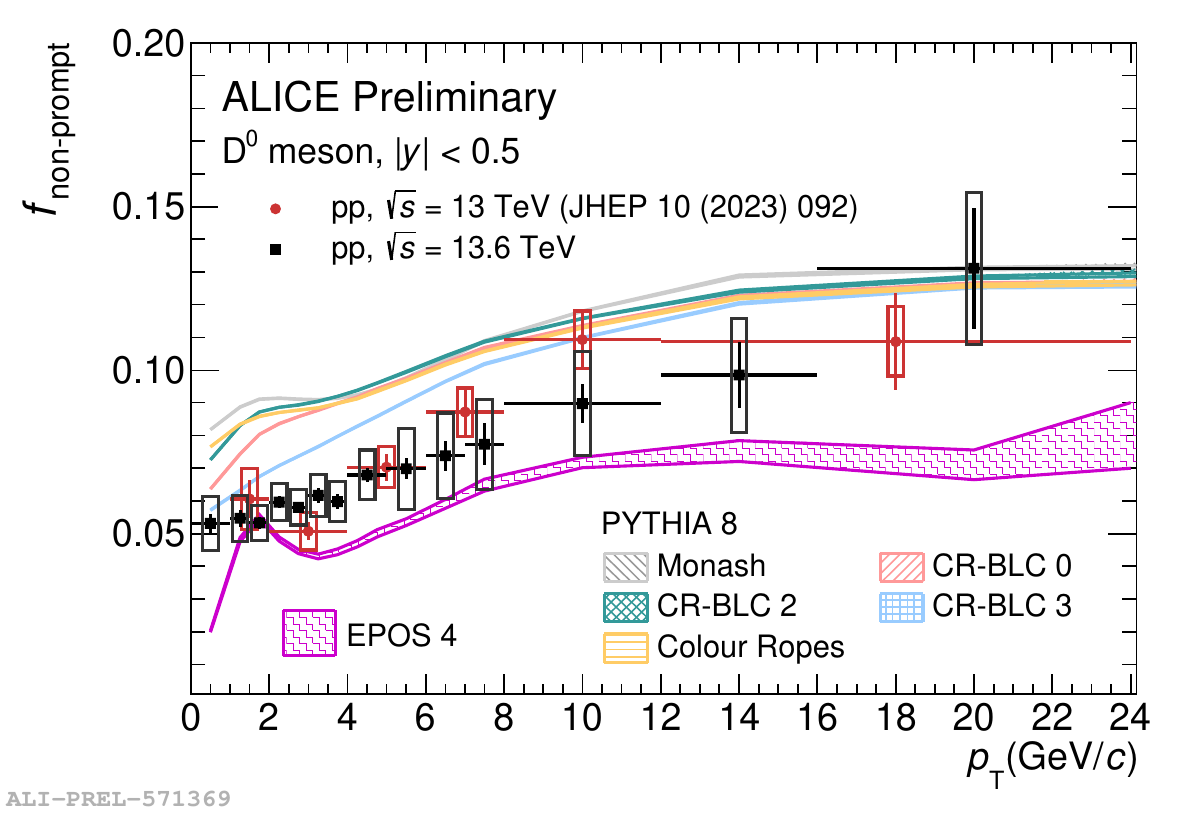}%
	}
	\caption{${\rm D}_{\rm s}^+/{\rm D}^+$ production (left), and
		non-prompt fraction of ${\rm D}^0$ production (right) at midrapidity ($|y|<0.5$) in pp collisions at $\sqrt{s}=13$ TeV compared with theoretical predictions.}
	\label{fig:Dratios}
\end{figure}

Contrary to charmed baryon-over-meson ratios, the new measurements in pp collisions at $\sqrt{s} = 13.6$ TeV from the 3${\rm rd}$ LHC data taking period (Run 3) show no strong \pt dependence in the strange to non-strange charmed-meson ratios ${\rm D}_{\rm s}^+ / {\rm D}^+$, as shown in Fig.~\ref{fig:Dratios} (left).
This trend is captured by the Catania model~\cite{Plumari:2017ntm}. However, it is overestimated by the POWLANG model~\cite{Beraudo:2015wsd}, and it is also not well reproduced by PYTHIA 8 simulations either with or without color junctions~\cite{Skands:2014pea,Christiansen:2015yqa}.

A fraction of charmed hadrons, the so-called non-prompt hadrons, come from decays of beauty hadrons. The non-prompt ${\rm D}^0$-meson fraction, as measured in  pp collisions by ALICE, is shown in Fig.~\ref{fig:Dratios} (right), together with theoretical calculations. Most of the calculations follow the trend correctly~\cite{Werner:2023zvo,Skands:2014pea,Christiansen:2015yqa,Bierlich:2014xba}, although the global scale is not reproduced well.

\begin{figure}[htb]
	\centerline{%
		\includegraphics[width=.9\columnwidth]{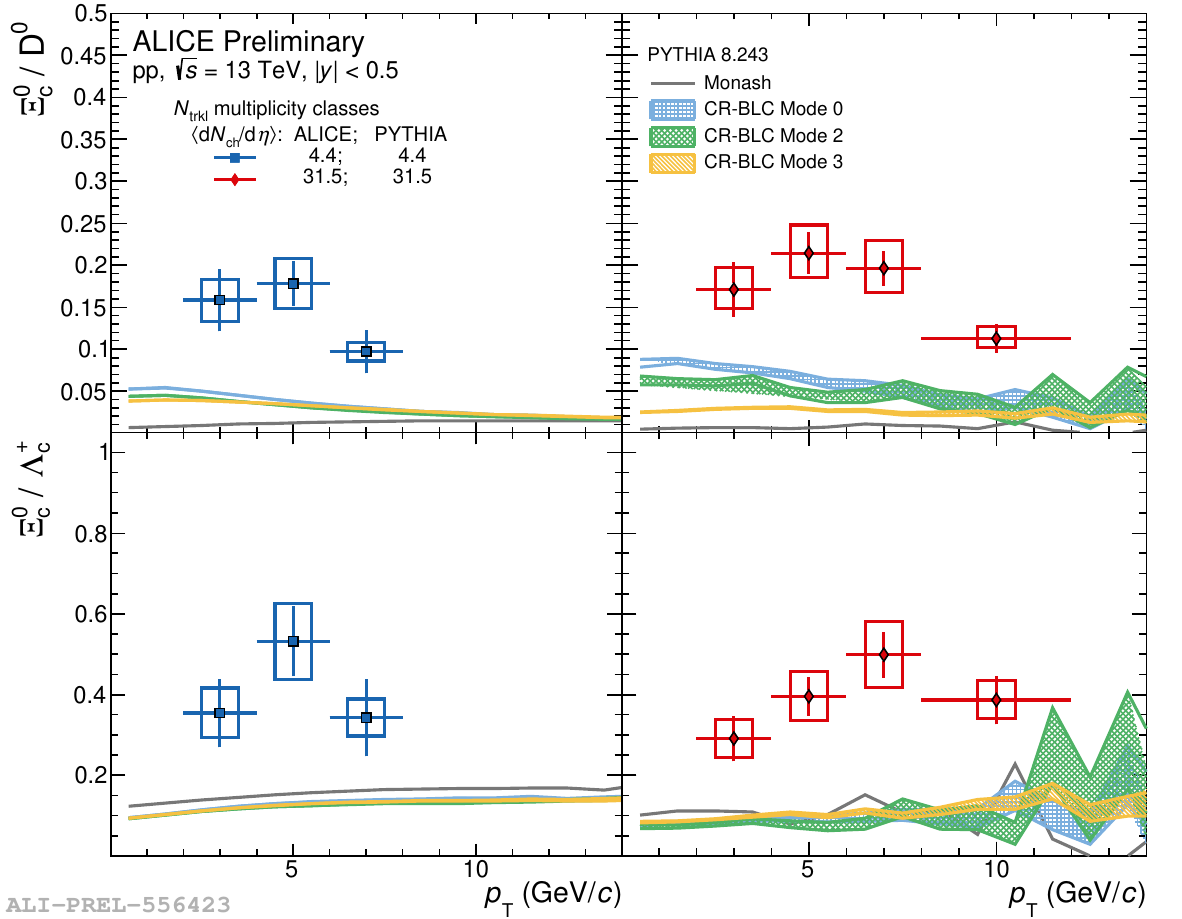}%
	}
	\caption{%
	The $\Xi_{\rm c}^0/{\rm D}^0$ (top) and $\Xi_{\rm c}^0/\Lambda_{\rm c}^+$ (bottom) ratios measured in pp collisions
	at $\sqrt{s}=13$ TeV for low (left) and high (right) multiplicity classes, at midrapidity ($|y|<0.5$). The results are compared to PYTHIA 8 results with and without color junctions.
	}
	\label{fig:Xic}
\end{figure}

Figure~\ref{fig:Xic} shows the ratio of the charmed-strange $\Xi_{\rm c}^0$ baryon production over that of charmed mesons and baryons, for different multiplicities.
Although no multiplicity dependence is seen within uncertainties, there is a hint that the results  depend on the \pt. The PYTHIA 8 model with color junctions~\cite{Christiansen:2015yqa} is unable to reproduce the observed ratio.

\subsection{Toward larger systems}

\begin{figure}[htb]
	\centerline{%
		\includegraphics[width=.49\columnwidth]{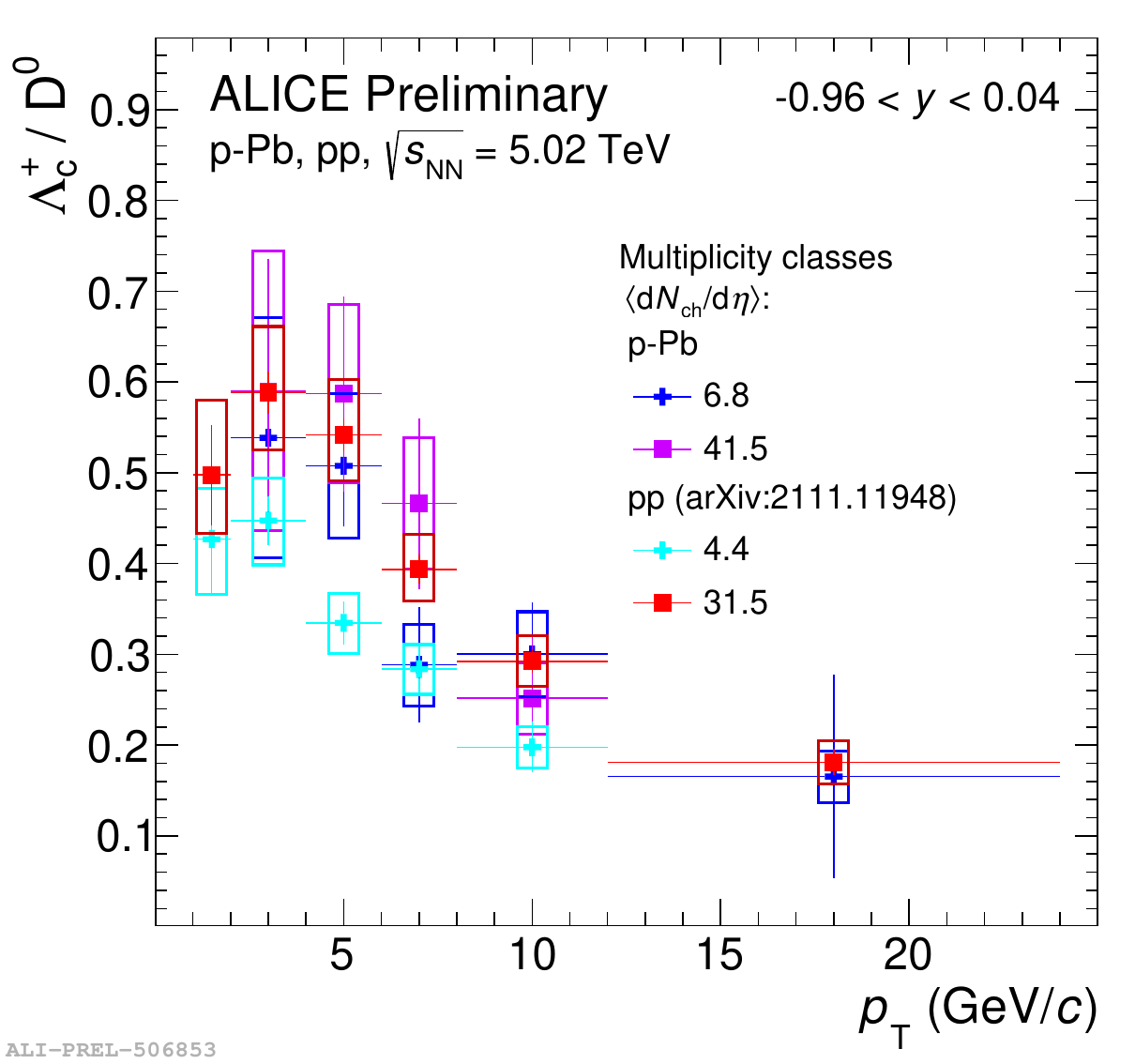}%
		\includegraphics[width=.49\columnwidth]{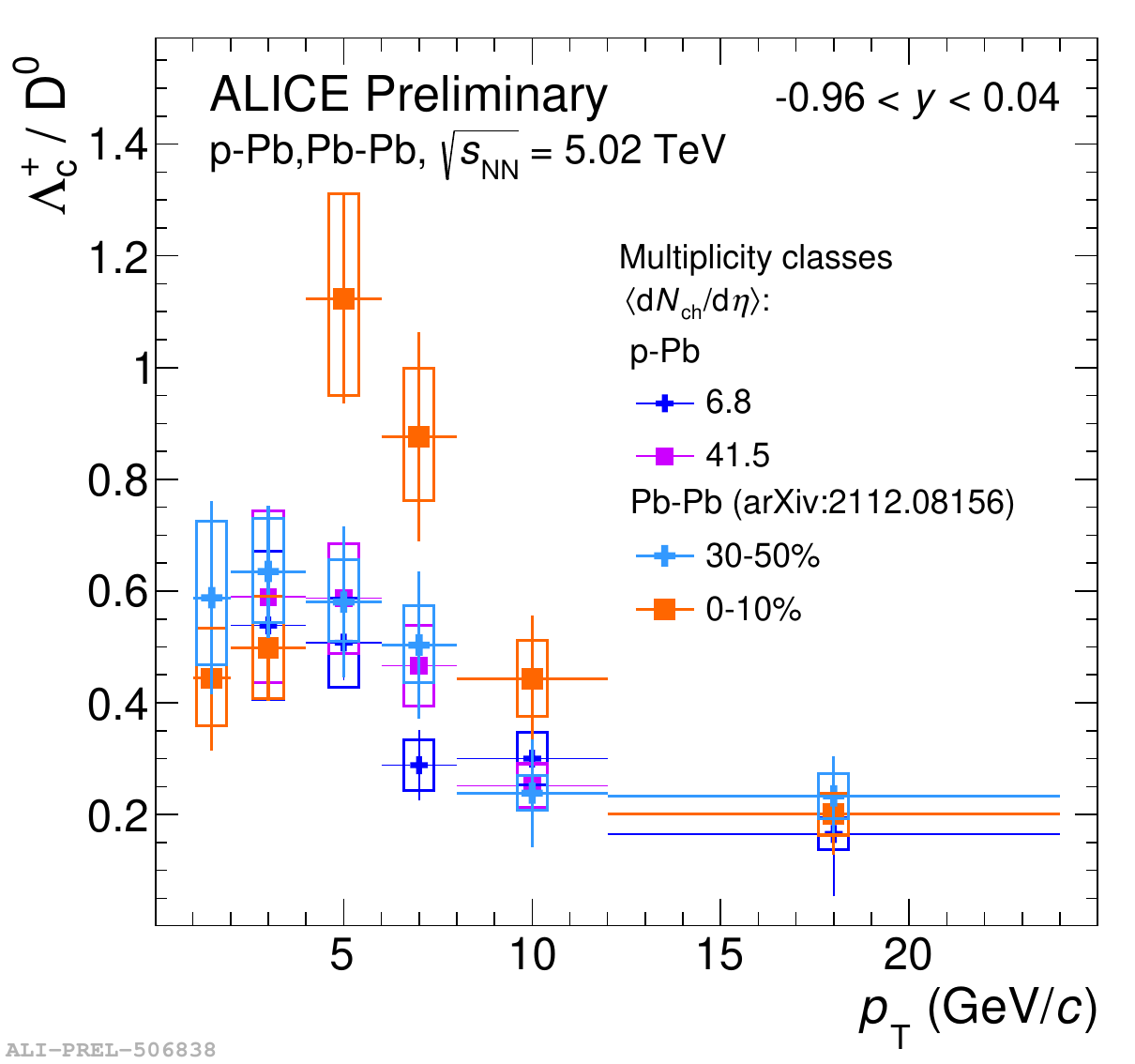}%
	}
	\caption{
		The $\Lambda_{\rm c}^+/{\rm D}^0$ ratio as a function of \pt in two multiplicity classes in the p--Pb collision at the $\sqrt{\it{s}_{\rm NN}}=5.02$ TeV, compared that in pp (left) and Pb--Pb (right) collisions.
	}
	\label{fig:LcDlargesyst}
\end{figure}

The phenomenon of charmed-baryon enhancement comes from multiple sources in large systems created in heavy-ion collisions. These may include high-multiplicity vacuum-QCD effects such as multiple-parton interactions with color reconnection (which in general produce events with high-mutiplicity charged particles), hot nuclear-matter effects such as collisional and radiative energy loss of heavy quarks, their participation in hydrodynamical evolution, thermalization and coalescence, as well as cold nuclear effects like shadowing. Comparative measurements of baryons, strange and non-strange
mesons from small to large collision systems can help clarify the picture.
Figure~\ref{fig:LcDlargesyst} (left) shows the $\Lambda_{\rm c}^+/{\rm D}^+$ ratios from pp and p--Pb collisions with different final-state charged-hadron multiplicities, while in Fig.~\ref{fig:LcDlargesyst} (right) the measurements from p--Pb are compared to semi-central and central Pb--Pb collisions.
An enhancement pattern similar to that in light baryon-to-meson ratios emerges at intermediate \pt. The high-multiplicity pp, low- and high-multiplicity p–Pb, and semi-central Pb--Pb collisions are consistent with each other, while the low-multiplicity pp measurements are separated. This suggests a threshold effect in the multiplicity-dependence of the $\Lambda_{\rm c}^+/{\rm D}^+$ ratios. 
A clearly stronger enhancement is present in central Pb--Pb collisions~\cite{ALICE:2021bib}. This trend is understood to result from the interplay between radial flow and coalescence~\cite{He:2019vgs,Plumari:2017ntm}.

Note that while the \pt-dependent trends of charmed-hadron production are sensitive to collision systems and different final-state multiplicities, there is no evidence of multi\-pli\-city dependence of the \pt-integrated charmed baryon-to-meson ratios from small to large hadron\-ic systems~\cite{ALICE:2021npz}.

\section{Summary and Outlook}

In summary, event-activity-dependent heavy-flavor measurements provide an opportunity to understand the complexity of pp collisions. Recent measurements have shown that hadronization is not universal across different collision systems, and explored the interplay between the effects in hot and cold nuclear matter and vacuum effects. Large systems can still be described within the standard picture of thermal equilibrium combined by hydrodynamical evolution.

While the first results from the LHC Run-3 data-taking period are ready, much more new data is expected in the near future. With the enhanced ITS and gas-electron-multiplier based TPC, two orders of magnitude enhancement can be expected in pp luminosity compared to Run-1 and Run-2 together. Precision and differential measurements and novel observables are becoming available, which can be used to disentangle possible sources of the observed effects.

This work has been supported by the NKFIH OTKA FK131979 and 2021-4.1.2-NEMZ\_KI-2024-00034 projects and the Wigner Scientific Computing Laboratory. 

\bibliographystyle{unsrt}
\bibliography{alice_hf_evtact-vertesi}

\end{document}